\documentclass[sigconf]{acmart}
\usepackage{booktabs}    
\usepackage{multirow}    
\usepackage{makecell}    
\usepackage{colortbl}    
\usepackage{xcolor}      
\usepackage{bm}
\usepackage{algorithm}
\usepackage[noend]{algpseudocode}

\AtBeginDocument{%
  }

\setcopyright{None}
\renewcommand\footnotetextcopyrightpermission[1]{}
\copyrightyear{2018}
\acmYear{2018}
\acmDOI{XXXXXXX.XXXXXXX}
\acmConference[Conference acronym 'XX]{Make sure to enter the correct
  conference title from your rights confirmation email}{June 03--05,
  2018}{Woodstock, NY}
\acmISBN{978-1-4503-XXXX-X/2018/06}

\begin{document}

\title{DesignCoder: Hierarchy-Aware and Self-Correcting UI Code Generation with Large Language Models}

\author{Yunnong Chen}
\affiliation{%
  \institution{College of Computer Science and Technology, Zhejiang University}
  \city{Hangzhou}
  \country{China}
}

\author{Shixian Ding}
\affiliation{%
  \institution{College of Computer Science and Technology, Zhejiang University}
  \city{Hangzhou}
  \country{China}
}

\author{Yinying Zhang}
\affiliation{%
  \institution{College of Computer Science and Technology, Jilin University}
  \city{Jilin}
  \country{China}
}

\author{Wenkai Chen}
\affiliation{%
  \institution{College of Computer Science and Technology, Zhejiang University}
  \city{Hangzhou}
  \country{China}
}

\author{Jinzhou Du}
\affiliation{%
  \institution{Huawei Technologies Co., Ltd.}
  \city{Shenzheng}
  \country{China}
}

\author{Lingyun Sun}
\affiliation{%
  \institution{College of Computer Science and Technology, Zhejiang University}
  \city{Hangzhou}
  \country{China}
}

\author{Liuqing Chen}
\affiliation{%
  \institution{College of Computer Science and Technology, Zhejiang University}
  \city{Hangzhou}
  \country{China}
}
\authornote{Corresponding author (chenlq@zju.edu.cn).}


\renewcommand{\shortauthors}{}

\begin{abstract}
Multimodal large language models (MLLMs) have streamlined front-end interface development by automating code generation. However, these models also introduce challenges in ensuring code quality. Existing approaches struggle to maintain both visual consistency and functional completeness in the generated components. Moreover, they lack mechanisms to assess the fidelity and correctness of the rendered pages. To address these issues, we propose DesignCoder, a novel hierarchical-aware and self-correcting automated code generation framework. Specifically, we introduce UI Grouping Chains, which enhance MLLMs' capability to understand and predict complex nested UI hierarchies. Subsequently, DesignCoder employs a hierarchical divide-and-conquer approach to generate front-end code. Finally, we incorporate a self-correction mechanism to improve the model’s ability to identify and rectify errors in the generated code. Extensive evaluations on a dataset of UI mockups collected from both open-source communities and industry projects demonstrate that DesignCoder outperforms state-of-the-art baselines in React Native, a widely adopted UI framework. Our method achieves a 37.63\%, 9.52\%, 12.82\% performance increase in visual similarity metrics (MSE, CLIP, SSIM) and significantly improves code structure similarity in terms of TreeBLEU, Container Match, and Tree Edit Distance by 30.19\%, 29.31\%, 24.67\%. Furthermore, we conducted a user study with professional developers to assess the quality and practicality of the generated code. Results indicate that DesignCoder aligns with industry best practices, demonstrating high usability, readability, and maintainability. Our approach provides an efficient and practical solution for agile front-end development, enabling development teams to focus more on core functionality and product innovation.
\end{abstract}



\maketitle
\section{Introduction}

GUI to code generation has impacted the front-end industry, accelerating development and reducing repetitive tasks for developers. Leveraging advances in deep learning and extensive open-source datasets, many efforts focus on translating visual representations into code, from sketches \cite{mohian2020doodle2app} and wireframes \cite{feng2023designing} to high-fidelity UI screenshots \cite{chen2018ui}. However, large-scale industrial development imposes strict requirements, including adherence to specific frameworks and the need for maintainable, accessible code. Several industry-level code generation platforms, such as Imgcook \cite{Imgcook} and CodeFun \cite{Codefun}, use mockups as input to generate code. These mockups created by Sketch \cite{Sketch} and Figma \cite{Figma} provide a visual representation and include essential design metadata, such as bounding boxes, hierarchy, and styles. Design inspection methods \cite{chen2023ui, chen2024egfe, chen2024fragmented} enhance code generation quality by detecting and resolving inconsistencies in design mockups. However, additional manual adjustments are still required to fully ensure the accuracy of the generated code.

\begin{figure}[t]
\centering
\includegraphics[width=0.48\textwidth]{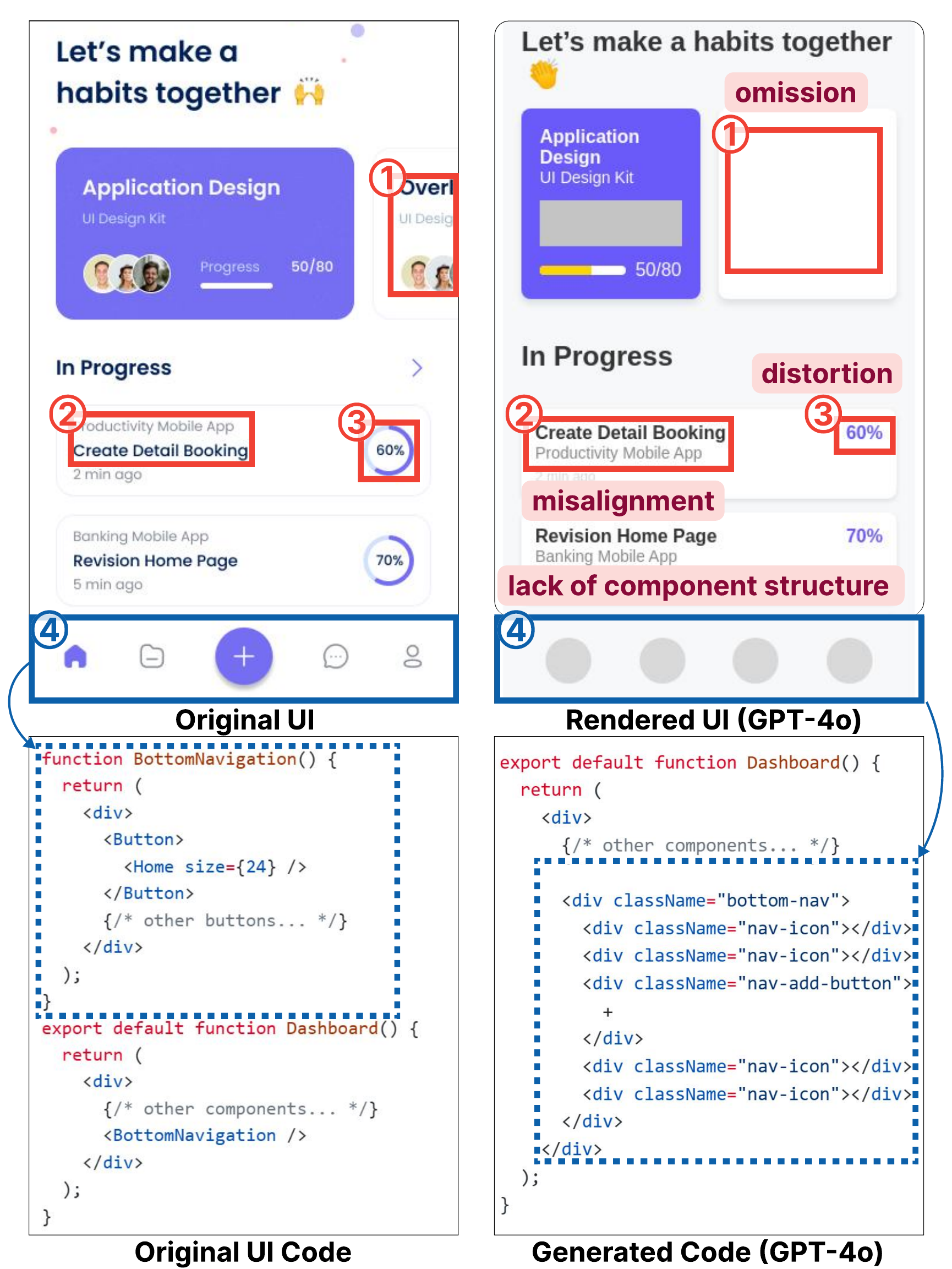}
\vspace{-0.15in}
\caption{Motivating examples are provided. Red boxes 1, 2, and 3 highlight element omission, misarrangement, and distortion in the rendered UI, respectively. Blue box 4 indicates that the generated code lacks a well-defined structure and clear functional semantics.}
\label{fig:motivation}
\vspace{-0.1in}
\end{figure}

Recent advancements in MLLMs have created new opportunities for GUI-to-code generation. Many studies have been proposed to enhance MLLMs’ code generation capabilities \cite{gui2025uicopilot, gui2025webcode2m, zhou2024bridging} or evaluate their performance \cite{si2024design2code,gui2024vision2ui}. Xiao et al. proposed Prototype2Code \cite{xiao2024prototype2code}, which combines element detection and layout information to generate HTML code, leveraging design metadata to prompt MLLMs for CSS code generation, showing promising results. DeclarUI \cite{zhou2024bridging} adopts computer vision models to generate function descriptions and introduce page transition graphs, guiding MLLMs to generate mobile app UI with jump logic. 

Despite these advancements, several major challenges remain. Existing MLLMs struggle to maintain visual consistency in generated code. For instance, the rendered output often exhibits missing elements, distorted components, and misaligned layouts, as highlighted by the red boxes in Figure \ref{fig:motivation}. This is because MLLMs lack a deep understanding of detailed UI styles, leading to lower fidelity in reconstruction. Additionally, MLLMs are prone to errors when generating lengthy code, which may result in missing elements and structural disorganization. Moreover, MLLMs face challenges in accurately interpreting the hierarchical structure of complex UI. While some methods \cite{xie2022psychologically,xiao2024ui} attempt to infer UI structures based on predefined rules, the rapid evolution of UI design reduces their effectiveness in adapting to new patterns. The blue boxes in Figure \ref{fig:motivation} illustrate an example of generated React code. We observe that correctly identifying the hierarchical structure of components not only enhances the modularity of the generated code but also improves component responsiveness. For example, buttons in the bottom navigation bar can be reused efficiently while adapting dynamically to different screen sizes. Additionally, traditional MLLMs lack program analysis capabilities and are unable to detect rendering errors, hindering the timely identification and correction of such issues.

To address these challenges, we propose DesignCoder, a novel hierarchy-aware, self-correcting GUI-to-Code generation method. To improve MLLM's ability to recognize hierarchical structures, we propose a UI grouping chain that integrates visual processing and grouping reasoning through a chain-of-thought process. The component tree, constructed through UI grouping chain, represents the hierarchical and semantic relationships of components within a UI mockup. We enrich this tree with style information extracted from UI mockup metadata. To generate structured component and style code, we propose a divide-and-conquer code generation strategy based on the component tree. After the initial code generation, we introduce vision-aware autonomous code repair to identify and correct issues. This multi-stage approach ensures that the generated UI code aligns closely with the original design in both visual fidelity and functional semantics.

To evaluate the effectiveness of DesignCoder, we collected a dataset of 300 high-fidelity mockups from the Figma community and a leading internet company. Under the React Native framework, our method outperforms the best baseline, achieving a 37.63\%, 9.52\%, 12.82\% improvement in visual similarity metrics (MSE, CLIP, SSIM) while significantly improving code structure similarity with increase of 30.19\%, 29.31\%, 24.67\% in TreeBLEU, Container Match, and Tree Edit Distance, respectively. Furthermore, we conducted a user study with professional developers to assess the quality and practicality of the generated code. The results indicate that DesignCoder significantly outperforms the baseline across multiple key aspects, including code availability (4.52 vs. 3.57), modification efficiency (4.20 vs. 3.24), readability (4.75 vs. 4.43), and maintainability (4.32 vs. 3.32). These findings highlight DesignCoder's capability to generate high-quality, production-ready front-end code at an industrial scale.

Our contributions are as follows:
\begin{itemize}
    \item We propose UI Grouping Chain, an innovation approach that leverages a UI multimodal chain-of-thought to guide MLLMs in UI layout recognition through visual processing and grouping reasoning.
    \item We introduce DesignCoder, a novel MLLM-based method for GUI-to-code generation, featuring hierarchy awareness and self-correction. This method generates structured components while leveraging design metadata, ensuring superior performance in both visual fidelity and code quality.
    \item We conduct a comprehensive evaluation of DesignCoder, comparing it with state-of-the-art baselines. Furthermore, we perform a user study, validating the effectiveness and practicality of our approach.
\end{itemize}
\section{Approach}
\begin{figure*}[thp]
\centering
\includegraphics[width=0.95\textwidth]{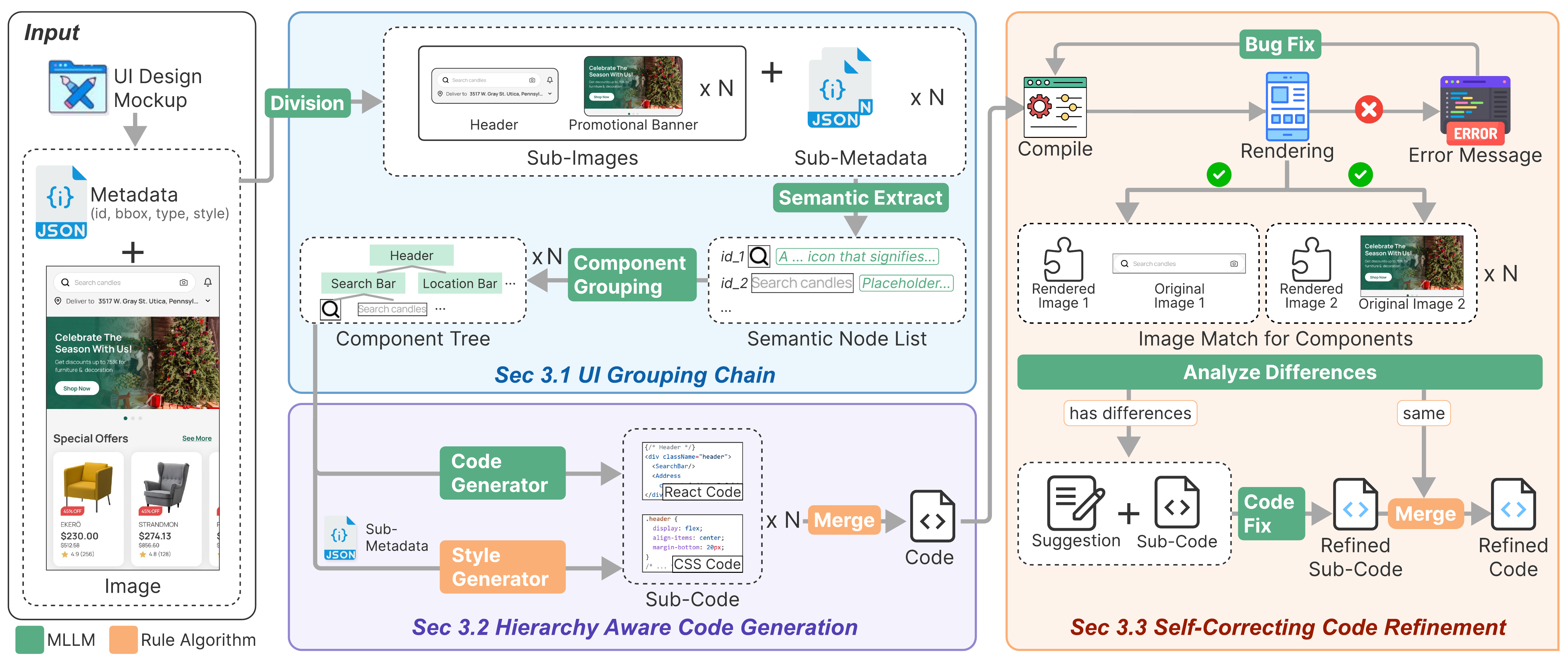}
\vspace{-0.1in}
\caption{The overview of DesignCoder.}
\label{fig:overview}
\vspace{-0.05in}
\end{figure*}

Figure \ref{fig:overview} illustrates the overall workflow of our method, DesignCoder, which takes UI design mockups as input. DesignCoder aims to accelerate development and iteration in real-world industrial scenarios, where visual consistency, code maintainability, and usability are critical. By leveraging design metadata, our method can acquire design information to enhance the quality of generated code. DesignCoder introduces the UI Grouping Chain (Section 3.1), which consists of two key steps: (1) Visual Processing: This step segments UI screenshots into subregions based on visual semantics. (2) Grouping Reasoning: MLLM employs a two-stage prompting strategy to analyze the semantics of elements within each subregion. It then integrates these semantics with visual prompts to guide the MLLM in constructing some hierarchical UI component sub-tree. In the code generation step (Section 3.2), we adopt a divide-and-conquer strategy, where each sub-tree is processed independently. The design metadata are used to generate CSS styles, and the layout and styles are integrated into a unified prompt to guide MLLM in generating code. Finally, in the self-correcting code refinement step (Section 3.3), our method iteratively verifies and corrects the initial generated code through visual comparison to ensure visual consistency and correctness.

\subsection{UI Grouping Chain}

\begin{figure}[thp]
\centering
\includegraphics[width=0.48\textwidth]{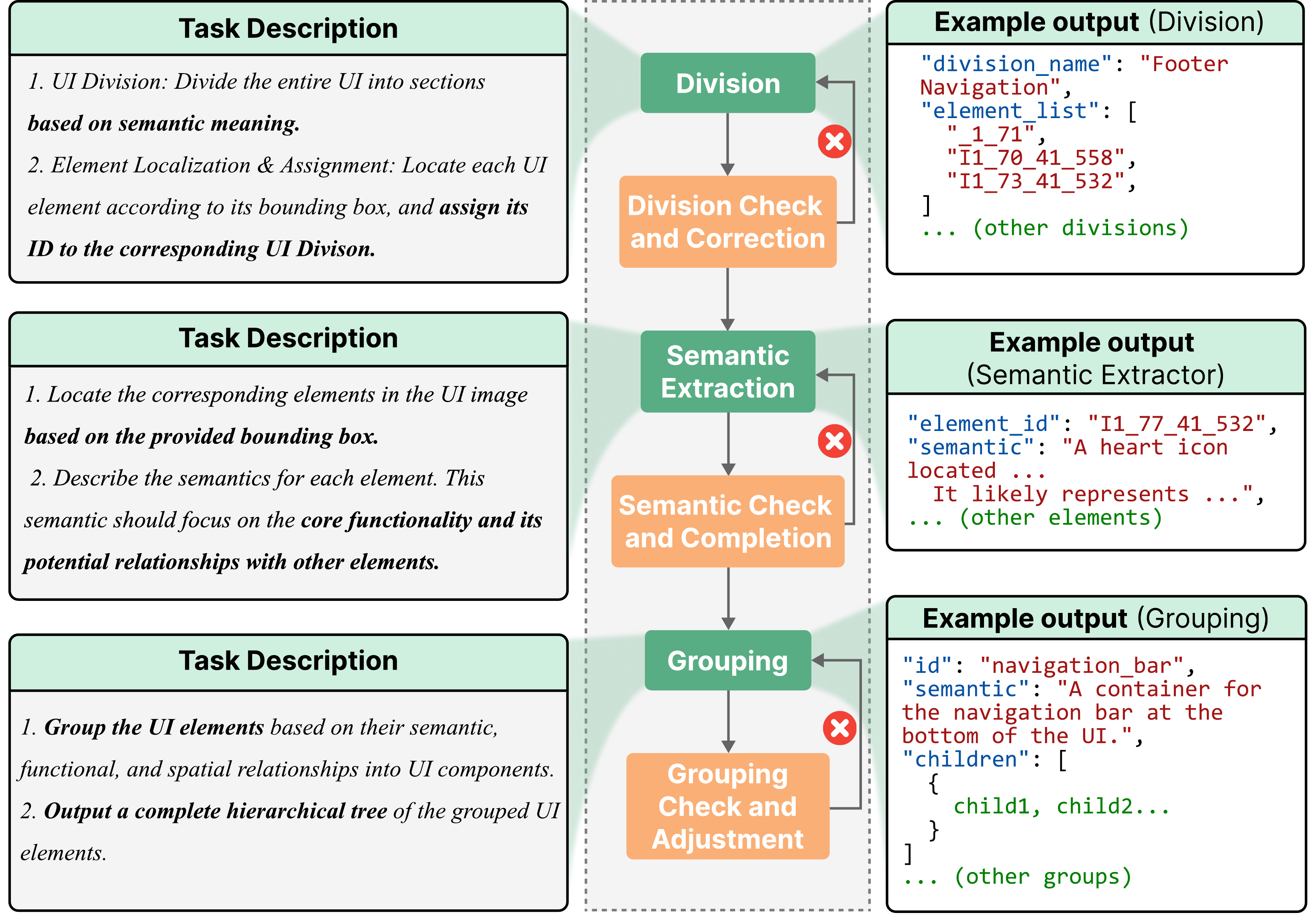}
\vspace{-0.15in}
\caption{The prompt used to construct UI grouping chain.}
\vspace{-0.1in}
\label{fig:prompt}
\end{figure}

Interpreting layer structures in design mockups and producing precise groupings is a challenging task for LLMs. To address this, we break down the grouping task into three more manageable subtasks, propose a UI grouping chain (Figure \ref{fig:prompt}), and integrate visual information to enhance the LLM’s effectiveness across these tasks. The three subtasks include UI division, semantic extraction, and component grouping. 

The UI comprises numerous elements, and the excessive layer information in design mockups poses challenges for MLLMs in processing. Prior research \cite{wan2024automatically} suggests that a divide-and-conquer strategy is an effective solution to this issue. By partitioning the UI into multiple regions and processing them separately, the work load on MLLMs can be significantly reduced. However, in mobile UI design, explicit boundaries are often absent, and UI regions are primarily defined based on semantic relationships. This constraint makes simple segmentation algorithms \cite{wan2024automatically} ineffective in accurately partitioning diverse mobile UIs. To address this challenge, we propose an LLM-based division tool that performs UI division by leveraging semantic relationships.

We first extract metadata from the design file, including id, type, and bbox, which serve to identify elements, provide semantic references, and determine element positions, respectively. This extracted information is then organized into a structured layer list \(L_{\text{layers}}\). Then, given the input image \(I\) and the \(L_{\text{layers}}\), we use the visual processing prompt \(P_{divide}\) to divide \(L_{\text{layers}}\) into multiple element sublists \(L_{\text{sub\_layers}}\). These element lists define each region’s position and dimensions. The entire UI is further partitioned based on this segmentation and proceeds to the next stage of processing. This process can be expressed as follows:
\begin{equation}
    L_{\text{layer}} = \big\{ (id_k, \mathbf{bbox}_k, \text{type}_k) \big\}_{k=1}^{N}
\end{equation}
\begin{equation}
    D = P_\text{divide}(I, L_{\text{layers}})
\end{equation}

To enhance the reliability of LLM-generated UI segmentation, validation and post-processing are essential. We establish three key principles: (1) each element belongs to a single region, except for large containers and backgrounds, which are handled separately; (2) regions must be mutually exclusive with no overlaps; (3) the number of regions should remain stable, with partitions outside the range of 3 to 10 requiring re-segmentation. As shown in Algorithm \ref{alg:postprocess}, by integrating an LLM with a division correction algorithm, our method effectively captures semantic and spatial information, ensuring high-precision segmentation.

\begin{algorithm}[t]
\caption{Check and Post-process}
\label{alg:postprocess}
\begin{algorithmic}
\Require Input Requirement ${D}$, $L_{layers}$
\Ensure Corrected divisions $\hat{D}$, where each layer is assigned to exactly one division.

\If{$|D| < 3 \vee |D| > 10 $}
    \State Rollback to $P_{divide}(I,L_{layers})$
\EndIf

\ForAll{$d \in D$}
    \State Compute bounding box $\mathrm{bbox}(d)$ by merging bboxes of $\forall{l} \in d_{layers}$
\EndFor

\ForAll{layers $l \in L_{layers}$}
    \State $I \gets \{\,d \in D \mid \mathrm{bbox}(l) \cap \mathrm{bbox}(d) \neq \emptyset\}$
    \If{$I = \emptyset$}
        \State Insert $l$ into the division $d^*$ whose center is closest to $d$
    \Else
        \State Merge all divisions in $I$ into a single division $d_m$ and insert $l$ into $d_m$
    \EndIf
    \State Update $\mathrm{bbox}$ of the affected division(s)
\EndFor

\ForAll{pairs of distinct divisions $(d_i, d_j)$}
    \If{$\mathrm{bbox}(d_i) \cap \mathrm{bbox}(d_j) \neq \emptyset$}
        \State Merge $d_i$ and $d_j$
    \EndIf
\EndFor

\State \textbf{return} Corrected divisions $\hat{D}$
\end{algorithmic}
\end{algorithm}

UI element grouping depends on both spatial and semantic relationships. To effectively incorporate the semantic information embedded in UI layers, we developed a tailored semantic extraction prompt \(P_{semantic}\) for UI layers. For each segmented subregion, we crop the corresponding UI image, extract relevant design metadata (i.e., type, bbox), and provide them with the prompt \(P_{semantic}\) as input to an MLLM. The model then produces an element list containing semantic annotations. For example, for text within a search bar, the generated description might be: \textit{"A placeholder text displaying 'Search for fruit salad combos,' which suggests the intended search scope or example use case for the search bar."} This process yields a layer semantic list \(L_\text{semantic}\). To refine the results, we apply a post-processing correction and employ visual enhancement techniques, such as annotating challenging-to-identify elements in the image, ensuring a comprehensive extraction of UI elements.

We introduce a structured grouping prompt \(P_{group}\) that organizes elements into a hierarchical component tree, using layer semantic list \(L_\text{semantic}\) and sub-images \(I_{sub\_image}\) as inputs. To ensure reliability, we implement a post-processing correction to prevent randomly generated leaf nodes and eliminates overlapping components. This structured approach transforms a flatten element list into a well-defined hierarchy, serving as the basis for front-end code generation. This process can be expressed as follows:
\begin{equation}
    L_{\text{semantic}} = P_{\text{semantic}}(L_{\text{sub\_layers}}, I_{\text{sub\_image}})
\end{equation}
\begin{equation}
    L_{\text{sub\_tree}} = P_{\text{group}}(L_{\text{semantic}}, I_{\text{sub\_image}})
\end{equation}

\subsection{Hierarchy-Aware Code Generation}

A complete declarative front-end implementation consists of two key steps: component code and CSS style code. For component code generation, we employ the UI hierarchical component tree generated by UI grouping chain as input. Our code prompt \(P_{code}\) comprises four steps: role assignment, task description, detailed requirement, and example output. The LLM is designated as a React expert and instructed to generate React components strictly following the component tree. It assigns subregion names as component identifiers, ensuring seamless association with style definitions and facilitating the assembly of the entire page. Leveraging its understanding of component semantics and interaction behaviors, the LLM selects appropriate component tags and assigns event handlers while leaving event content unspecified, facilitating further application development. For example, in the case of scrollable views, if the design explicitly defines a scrolling behavior, our model can intelligently select the <ScrollView> component rather than simply applying overflow: scroll. This approach enhances both code readability and maintainability. This process can be expressed as follows:

\begin{equation}
    L_{\text{code}} = P_{\text{code}}(L_{\text{sub\_tree}}, I_{\text{sub\_image}})
\end{equation}

CSS provides visual styles such as colors, layout, borders, and spacing. To maintain visual consistency between the generated code and the original UI, we utilize design metadata and structured component information. For layout generation, we traverse the hierarchy (layer) tree from bottom to top. Each leaf node corresponds to an element and retains its original bbox attributes, while non-leaf nodes aggregate bounding box data from child nodes to determine size and relative positioning. This aggregation helps ensure accurate alignment and visual consistency. Attributes for text (such as style, color, font, and weight) are systematically extracted to compute relative sizes, spacing, and line heights. Additionally, padding, borders, shadows, and corner styles are generated according to the design metadata. Finally, the structured front-end code is organized into subregions, accurately reflecting UI layout properties, providing responsiveness across different screen sizes.

\subsection{Self-correcting Code Refinement}

Since LLMs are generated in an auto-regressive way, once an error occurs during the generation process, the LLMs will continue to generate content on the basis of the errors, leading to the propagation of errors. The code generated by MLLMs may contain various errors, including missing elements, style distortions, layout inconsistencies, syntax errors, and functional anomalies. While existing studies \cite{zhou2024bridging, jiang2024rocode} focus on ensuring correct code parsing and execution, correct parsing does not necessarily guarantee that the generated code is visually and functionally accurate. For instance, the generated code may omit the search icon within the search bar or result in a disorganized element layout, ultimately affecting usability. To address these issues, we propose a self-correcting strategy based on the generated component tree, which systematically evaluates the visual consistency and responsive layout of the generated code. We then employ a divide-and-conquer approach to perform fine-grained code corrections and merge the refined code.

Specifically, we leverage Appium—a UI automation testing tool—to capture rendered page screenshots and extract component attributes, including bounding boxes, element types, text content, and hierarchical structures. Next, we match the rendered components using component IDs from the component tree. Then, we extract image segments based on each component’s bounding box, obtaining the original component image \( I_j \) and the rendered component image \( I_k \).

We then employ a visual analysis prompt, \( P_{\text{analysis}} \), to meticulously compare these images, focusing on identifying any element misarrangement or style errors in the generated code:
\begin{equation}
S_i = P_{\text{analysis}}(I_j, I_k)   
\end{equation}

where \( S_i \) represents the repair suggestion for the \( i \)-th component, with \( I_j \) corresponding to the original image and \( I_k \) to the rendered image.

Subsequently, the generated repair suggestion \( S_i \) is combined with the associated component code snippet to form a structured code repair prompt \( P_{\text{repair}} \):
\begin{equation}
P_{\text{repair}} = \{ \text{``component\_code\_i''},\ S_i \}
\end{equation}

\begin{figure}[thp]
\centering
\includegraphics[width=0.46\textwidth]{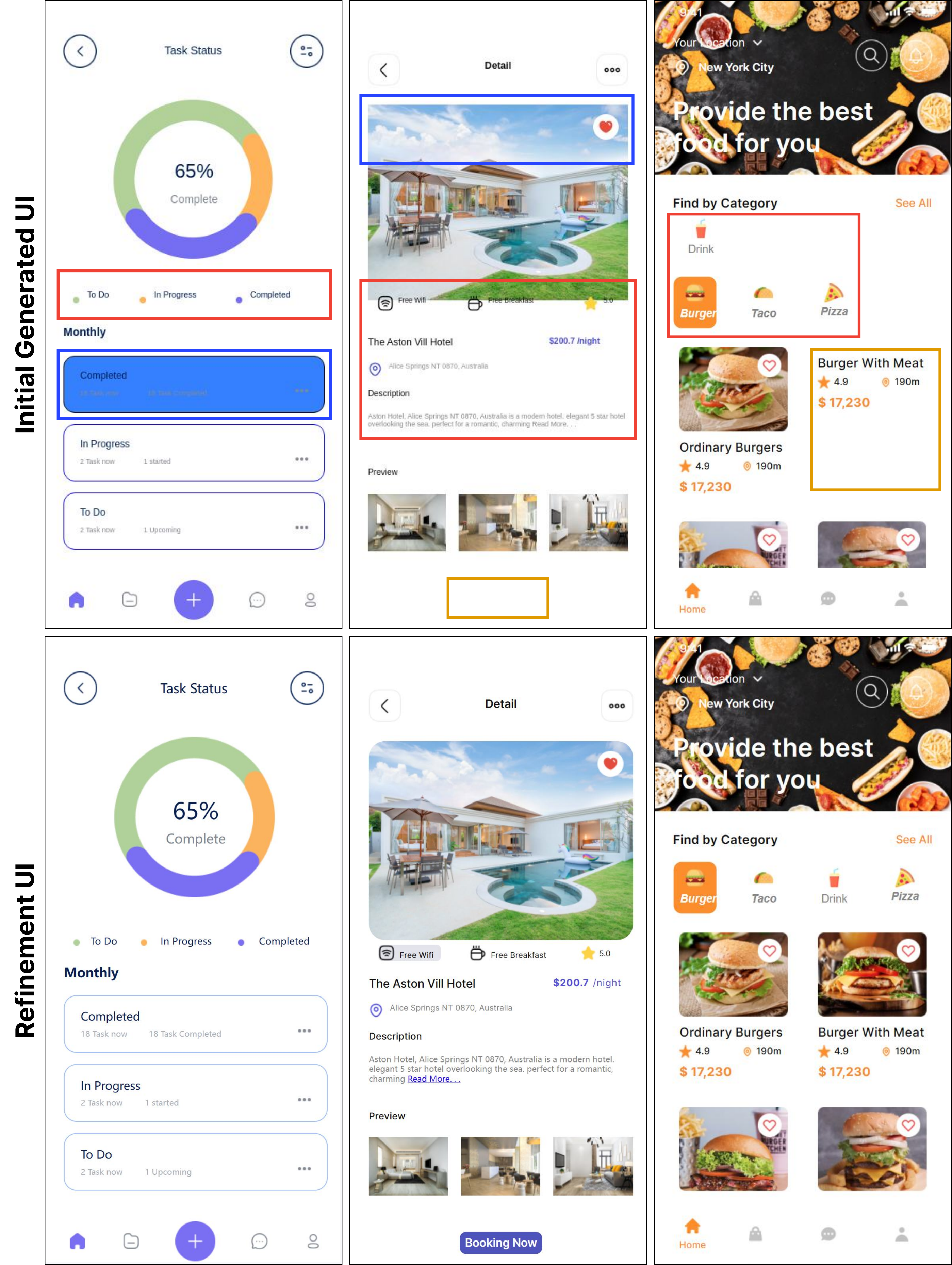}
\vspace{-0.1in}
\caption{Examples of the rendering results from both the initially generated code and the refinement code. The red boxes highlight misaligned elements, the blue boxes indicate visual distortions, and the yellow boxes mark missing elements.}
\vspace{-0.1in}
\label{fig:fix}
\end{figure}

This prompt guides the MLLM in performing targeted code repairs. Notably, since each component can be repaired independently, the process can be parallelized, greatly enhancing the efficiency of code repair. Finally, we merge the code snippets according to the component tree order to produce the final rendered page. Figure \ref{fig:fix} presents three examples, including both the initially generated code and its corresponding refined code. We identified three typical types of errors, which are highlighted in the Figure \ref{fig:fix}: red boxes indicate misaligned elements, blue boxes represent distorted elements, and yellow boxes denote missing elements. By comparing the rendered results with the original UI, we found that MLLM effectively detects potential errors and accurately locates the corresponding code segments. For instance, in Case 1, the blue box marks an incorrectly generated blue background, while the red box highlights misaligned elements. Our method successfully detected these errors and applied the necessary corrections. In Case 2, our approach not only identified overlapping elements (marked in the red box) but also adjusted the rounded corners of an background image (marked in the blue box), demonstrating its ability to perform fine-grained code corrections. This improvement stems from our strategy of breaking down the code repair task based on the generated component tree, employing a divide-and-conquer approach to guide MLLM in the correction process. These results suggest that improving MLLMs’ understanding of errors in generated code is a key factor in enhancing code generation quality.

\section{Evaluation}
To comprehensively evaluate the performance of DesignCoder, we designed a series of experiments to address the following three key research questions (RQs):
\begin{itemize}
    \item  \textbf{RQ1: Effectiveness.} How does DesignCoder perform in terms of visual consistency and structural similarity compared to baseline methods?
    \item \textbf{RQ2: Ablation Study.} What is the contribution of each key component to the overall performance of DesignCoder?
    \item \textbf{RQ3: User Study and Case Study.} What is the perceived quality of the generated code by DesignCoder from a user perspective?
\end{itemize}

\subsection{Dataset}
We constructed a dataset containing 300 mobile application UI design mockups, spanning five categories: entertainment, education, health, shopping, and travel. The dataset was sourced from the Figma community and a large internet company using Sketch software. Specifically, it includes 250 mockups from the Figma community and 50 from the enterprise source. The Sketch files were imported into Figma, and we employed the Figma CLI to extract standardized image assets and design metadata. Importantly, since the hierarchical structures of design mockups do not directly correspond to code structures, our method does not depend on the original layer structures.

\subsection{Experimental Setup}
\subsubsection{Baselines.}
In this study, our baseline methods can be categorized into two types: methods that are specialized for mobile UI generation and general-purpose MLLMs. The specialized methods include:
\begin{itemize}
    \item \textbf{Prototype2Code} \cite{xiao2024prototype2code} enhances the hierarchical structure of design mockups by utilizing fragmented elements grouping \cite{chen2024egfe} and semantic grouping techniques \cite{xie2022psychologically}. It employs hand-crafted rules to generate a layout tree, recognize element types for constructing the HTML skeleton code, and then leverage a large language model to generate CSS style code.
    \item \textbf{DeclarUI} \cite{zhou2024bridging} integrates an unsupervised element detection model \cite{liu2024grounding} and segment anything model \cite{kirillov2023segment} to generate UI component functional description. It then introduce page transition graphs to build interactive logic, which are used to prompt MLLMs for generating mobile app UIs with jump logic. We use its GPT-4o version as the baseline.
\end{itemize}

The general-purpose MLLMs include:
\begin{itemize}
    \item \textbf{GPT-4o}, developed by OpenAI, exhibits exceptional image understanding capabilities and can generate code based on images. Following the best practices in existing image-to-code research \cite{si2024design2code, wu2024uicoder, xiao2024prototype2code}, we have meticulously designed the prompts for GPT-4o (specifically, GPT-4o-2024-11-20).
    \item \textbf{Claude-3.5}, developed by Anthropic, is a multimodal large model that has demonstrated outstanding performance across multiple MLLM benchmarks. We use Claude-3.5-Sonnet-20241022 as the baseline.
    \item \textbf{LLaVA-v1.5-7B} \cite{liu2023visual} is an end-to-end trained multimodal large model that enables general vision-language understanding by integrating a vision encoder with a large language model. We use the same prompts as GPT-4o to generate UI code from images.
\end{itemize}

\subsubsection{Evaluation Metrics}
Our evaluation employs a set of metrics to assess the performance of DesignCoder, encompassing visual similarity, structure similarity, and practical usability.

To evaluate visual similarity, we use UI screenshots extracted from UI mockups as the ground truth and introduce three metrics:
\begin{itemize}
    \item \textbf{MSE} is a common metric for evaluating image similarity by measuring the pixel-wise differences between two images. It calculates the average squared difference between corresponding pixel values, where lower values indicate higher similarity.
    \item \textbf{CLIP Score} \cite{si2024design2code}. This metric quantifies the semantic similarity between the generated and original UIs, defined by $\cos\left(\operatorname{clip}(\text{image}_i), \operatorname{clip}(\text{image}_j)\right)$, which measures the cosine similarity between the CLIP feature representations of the images.
    \item \textbf{Structural Similarity Index Measure (SSIM)} \cite{wang2004image}. It assesses the layout and compositional accuracy, focusing on the spatial arrangement similarities between the generated and original UIs. A limitation of SSIM is that it only evaluates structural similarity at the vision level without assessing structural and semantic similarity at the code level. 
\end{itemize}

Therefore, to evaluate the structural similarity of the generated code, we adopted a combination of automated and manual code implementation. To obtain the ground truth of UI code, we introduce Codefun \cite{Codefun}, an enterprise-level code generation platform, which converts well-structured design files into front-end code. Subsequently, two authors manually inspected the generated component tree, focusing on hierarchy and component tags. For cases with structural hierarchy errors, each author independently corrected the issues before reaching a consensus through discussion. Additionally, the authors referred to Material Design 3 to rectify incorrect component types. For example, a text button labeled as a ``text'' type was corrected to a ``button'' type. Then, we introduce three metrics to compare hierarchical structures:

\begin{itemize}
    \item \textbf{TreeBLEU} \cite{gui2025webcode2m} is used to evaluate the similarity between the DOM tree of the generated HTML and the ground truth, excluding terminal nodes that contain tag attributes (e.g., content and styles). It is defined as the proportion of all height-1 subtrees in the given tree that match subtrees in the reference tree. TreeBLEU provides a holistic measure of HTML DOM tree similarity. We have adapted this metric to enhance its applicability to mobile code evaluation. It can  be formulated as:
    \begin{equation}
        \text{TreeBLEU} = \frac{|S(t) \cap S(t_r)|}{|S(t_r)|},
    \end{equation}
    where \( t \) and \( t_r \) denote the given tree and the reference tree, respectively.
    \item \textbf{Tree Edit Distance} (TED) is a metric for measuring tree similarity based on the minimum number of ``edit'' operations required to transform one tree into another isomorphic tree. 
    \begin{equation}
    \text{TED}(t_1, t_2) = \min_{(e_1, \dots, e_k) \in \mathcal{P}(t_1, t_2)} \sum_{i=1}^{k} c(e_i)
    \end{equation}
    \(\mathcal{P}(t_1, t_2)\) refers to the set of possible edit paths between \( t_1 \) and \( t_2 \).
    A lower TED indicates better performance. Therefore, if there are no matching nodes in the predicted result, we set TED to the number of edges in the ground truth tree.
    \item \textbf{Container Match} (CM) \cite{wu2021screen} focuses on the grouping of elements rather than their specific structure. This metric is particularly relevant for certain downstream tasks, such as screen segmentation, where the interface is divided into distinct regions. The metric is computed by averaging the Intersection over Union (IoU) scores of each container (e.g., intermediate nodes) to the ground truth annotations. The Container Match score ranges from 0 to 1, where CM = 1 indicates perfect grouping alignment. For tree structures with no matching nodes, we assign a score of 0.
\end{itemize}

\subsubsection{Implementation Details}
In alignment with DeclarUI, we selected GPT-4o as the foundational model. In our experiments, all generated code was implemented using popular mobile development frameworks, specifically React Native \cite{react}. We automated the integration of the generated code into an initial project within Android Studio using Python scripts to minimize human intervention and bias. For code that failed to compile, we assigned a value of zero to all its evaluation metrics.

\subsection{RQ1: Effectiveness}
\textbf{Visual metric.} Table \ref{tb_results_1} indicates that DesignCoder outperforms all baseline methods regarding the visual metrics. Compared to the state-of-the-art baseline Prototype2Code, our approach achieves improvements of 37.64\%, 9.52\%, and 12.82\% in MSE, CLIP, and SSIM scores on the Figma dataset, and 24.31\%, 4.60\%, and 11.69\% on the company dataset. The SSIM and CLIP scores indicate that, in terms of both image structure and semantics, the code generated by DesignCoder aligns closely with human perceptual judgments. For MSE score, a lower value indicates better visual consistency, suggesting that DesignCoder generates UI code that more accurately reproduces the original design. This is because our method captures the hierarchical relationships of UI components and incorporates design metadata, leading to improved visual fidelity. Compared to the DeclarUI, which relies solely on image input, our method outperforms it when using the same foundation model (GPT-4o). This is because, although using CV models can enhance MLLMs' understanding of UI images, the complex hierarchical relationships in UI and the large-scale variations in UI components (such as background images and icons) pose significant challenges to the generalization capability of CV models. Our method leverages UI multimodal chain-of-thought to enhance MLLMs’ capability in UI understanding, enabling a deeper comprehension of the semantic relationships among UI components. Furthermore, we introduce a self-correcting code refinement mechanism, allowing the LLM to detect subtle issues and autonomously correct them, thereby achieving greater visual consistency with the original UI. 

\textbf{Structural metric.} We have found that visual metrics alone cannot fully reflect the quality of generated code. In actual development, front-end developers place greater emphasis on whether the code adheres to good structural norms. Well-structured code is generally more maintainable and easier to modify. Therefore, we introduced TreeBLEU, Tree Edit Distance, and Container Match to assess code structure. 

The results in Table \ref{tb_results_2} show that, compared to the best baseline, our method improves TreeBLEU, Container Match, and Tree Edit Distance on the Figma dataset by 30.19\%, 29.31\%, and 29.47\%, on the company dataset by 17.78\%, 22.45\%, and 26.33\%, respectively. This indicates that our approach better captures common component types and hierarchical relationships. It is worth noting that although DeclarUI considers page navigation logic, it overlooks intra-page component relationships, which affects the generated code structure. More importantly, in the Company Dataset, which features more complex component relationships, the code structure generated by DesignCoder is also better organized. This further demonstrates DesignCoder’s strong generalization ability and adaptability, benefiting from the extensive training data and large-scale parameters of MLLMs.

\begin{table}[htp]
\vspace{-2mm}
\centering
\caption{{Comparison of Visual Similarity Metrics Across Different Methods (the best is marked in bold).}}
\vspace{-2mm}
\resizebox{1\columnwidth}{!}{%
\begin{tabular}{lcccccc}
\toprule[1.2pt]
& \multicolumn{3}{c}{\textbf{Figma Dataset}} 
& \multicolumn{3}{c}{\textbf{Company Dataset}} \\
\cmidrule(lr){2-4} \cmidrule(lr){5-7}
\textbf{Model} & \textbf{MSE}$\downarrow$ & \textbf{CLIP}$\uparrow$ & \textbf{SSIM}$\uparrow$ & \textbf{MSE}$\downarrow$ & \textbf{CLIP}$\uparrow$ & \textbf{SSIM}$\uparrow$ \\
\midrule
LLava-v1.5-7B     & 91.49 & 0.58 & 0.58 & 85.28 & 0.57 & 0.56 \\
GPT-4o            & 70.59 & 0.69 & 0.61 & 72.13 & 0.61 & 0.69 \\
Claude-3.5        & 82.47 & 0.67 & 0.52 & 83.62 & 0.62 & 0.55 \\
DeclarUI          & 52.83 & 0.79 & 0.83 & 53.76 & 0.75 & 0.70 \\
Prototype2Code    & 36.32 & 0.84 & 0.78 & 28.18 & 0.87 & 0.77 \\
CodeFun           & 28.45 & \textbf{0.93} & 0.71 & 58.55 & \textbf{0.91} & 0.57 \\
\rowcolor{gray!10!white}DesignCoder 
& \textbf{22.65} & 0.92 & \textbf{0.88} & \textbf{21.33} & \textbf{0.91} & \textbf{0.86} \\
\bottomrule[1.2pt]
\end{tabular}
}
\label{tb_results_1}
\vspace{-2mm}
\end{table}

\begin{table*}[htp]
\vspace{-2mm}
\centering
\caption{{Comparison of Structural Similarity Metrics Across Different Methods (the best is marked in bold).}}
\vspace{-2mm}
\resizebox{0.95\textwidth}{!}{%
\begin{tabular}{lcccccc}
\toprule[1.2pt]
& \multicolumn{3}{c}{\textbf{Figma Dataset}} 
& \multicolumn{3}{c}{\textbf{Company Dataset}} \\
\cmidrule(lr){2-4} \cmidrule(lr){5-7}
\textbf{Model} & \textbf{TreeBLEU}$\uparrow$ &  \textbf{Container Match}$\uparrow$ & \textbf{Tree Edit Distance}$\downarrow$ &\textbf{TreeBLEU}$\uparrow$ & \textbf{Container Match}$\uparrow$ & \textbf{Tree Edit Distance}$\downarrow$ \\
\midrule
LLava-v1.5-7B     & 0.19 & 0.25 & 49.11 & 0.11 & 0.18 & 42.54 \\
GPT-4o            & 0.24 & 0.31 & 49.85 & 0.22 & 0.30 & 32.77 \\
Claude-3.5        & 0.24 & 0.33 & 41.49 & 0.19 & 0.27 & 35.62 \\
DeclarUI          & 0.32 & 0.41 & 43.12 & 0.28 & 0.34 & 31.10 \\
Prototype2Code    & 0.53 & 0.58 & 40.00 & 0.45 & 0.49 & 28.50 \\
CodeFun           & 0.52 & 0.61 & 35.63 & 0.41 & 0.52 & 30.45 \\
\rowcolor{gray!10!white}DesignCoder 
& \textbf{0.69} & \textbf{0.75} & \textbf{28.21} 
& \textbf{0.53} & \textbf{0.60} & \textbf{22.86} \\
\bottomrule[1.2pt]
\end{tabular}
}
\label{tb_results_2}
\vspace{-2mm}
\end{table*}

\subsection{RQ2: Ablation Study}
We conducted a ablation study to investigate the effectiveness of the key components in DesignCoder. The results of these ablation studies are presented in Table \ref{tab:ablation}. In the first ablation experiment, we removed the UI Grouping Chain module. Instead, we directly prompted MLLMs to generate the component tree using UI screenshots and a list of layers as input. Removing the UI Grouping Chain module led to a significant decline in visual similarity scores. Specifically, the CLIP score dropped from 0.92 to 0.84, the SSIM decreased from 0.88 to 0.79, and the MSE increased from 22.43 to 38.92. These declines in visual metrics directly indicate a reduced ability of the MLLM to generate structurally correct component trees. This finding highlights the challenge of allowing MLLM to process complex UI pages directly. Our method helps MLLM break down complex tasks into simpler subtasks by decomposing UI images into multiple sub-images, thereby reducing cognitive load. Moreover, the code structure similarity scores also dropped significantly, with TreeBLEU decreasing by 42.4\% and Container Match by 43.8\%. This decline is likely due to the absence of component-level semantic reasoning in MLLM. Semantic relationships enhance MLLM’s ability to infer grouping logic, enabling the recognition of complex hierarchical structures.

In the second ablation experiment, we removed the code refinement module. As shown in Table \ref{tab:ablation}, both visual and code structure similarity scores declined. We observed that MLLM can identify certain detailed style issues, such as font size, missing or misarranged elements, and color distortion, and effectively correct them in the code. Additionally, visual comparison can reveal code structure issues, such as missing containers that lead to incorrect relative positioning. Our method is capable of detecting these issues and making timely corrections.

\begin{table*}[htp]
    \centering
    \caption{Comprehensive Performance Metrics for Ablation Study (merged two datasets, with the best is marked in bold).}
    \vspace{-2mm}
    \resizebox{0.75\textwidth}{!}{%
    \begin{tabular}{lcccccc}
    \toprule[1.2pt]
        \textbf{Method} & \textbf{MSE}$\downarrow$ & \textbf{CLIP}$\uparrow$ & \textbf{SSIM}$\uparrow$ & \textbf{TreeBLEU}$\uparrow$ & \textbf{Container Match}$\uparrow$ & \textbf{Tree Edit Distace}$\downarrow$  \\
        \hline
        \rowcolor{gray!10!white}DesignCoder & 22.43 & 0.92 & 0.88 & 0.66 & 0.73 & 27.32\\
        w/o Grouping & 38.92 & 0.84 &0.79 & 0.38 & 0.41 & 36.57\\
        w/o Refinement & 33.64 & 0.88 & 0.85 & 0.55 & 0.64 & 29.44 \\
    \bottomrule[1.2pt]
    \end{tabular}
    }
    \label{tab:ablation}
\end{table*}

\subsection{RQ3: User Study and Case Study}

\subsubsection{User Study}
In large-scale industrial development, the overall quality of code is a key criterion for evaluating the performance of automatic code generation methods \cite{chen2024egfe,xiao2024prototype2code,zhou2024bridging}. To this end, we conducted a user study, inviting experienced developers to assess the generated code and participate in interviews to discuss the rationale behind their evaluations.

\textbf{Procedures.} We recruited five participants with over five years of experience in front-end development using the React Native framework. These participants can be considered senior front-end developers capable of evaluating the quality of the generated code. To conduct our study, we randomly selected 15 UI design prototypes representing three typical UI design categories: travel, communication, and shopping. Among these, 10 prototypes were sourced from Figma, while the remaining five were obtained from the company. We used DeclarUI, equipped with GPT-4o, as the baseline for comparison. Each participant independently reviewed the code generated by both DesignCoder and the baseline. Following \cite{chen2024egfe}, we assessed the modification process based on two key metrics: code availability and modification time. After completing the modifications, participants rated the readability and maintainability of the code using a five-point Likert scale. Code availability was tracked via Git logs, while modification time measured the duration required to adjust the code to production standards. The time was categorized into five-minute intervals, with a score of 5 indicating completion within five minutes and a score of 1 indicating completion exceeding 20 minutes. All participants conducted the evaluations using Android Studio.

\textbf{Results.} Table \ref{tab:human evaluation} presents the average scores, with p-values reported using the Mann-Whitney U test. Overall, DesignCoder outperforms the Baseline across four code quality evaluation metrics: code availability, modification time, readability, and maintainability, with average scores of 4.52, 4.20, 4.75, and 4.32, respectively. Notably, our approach shows significant differences from DeclarUI across all four metrics. The higher maintainability and readability scores indicate that the code generated by DesignCoder is easier to maintain in production environments. Participants consistently agreed that DesignCoder’s generated code exhibits better responsive design and modularity. This is attributed to the well-structured hierarchy among components and DesignCoder’s ability to correctly identify relationships between them. Participant P1 commented: \textit{“Compared to the LLMs I have used before, the structure (of the generated code) is much closer to what I write in my projects.”} Additionally, DesignCoder significantly reduces code modification time (DesignCoder: 4.20 vs. Baseline: 3.24), suggesting that DesignCoder-generated code is easier to adjust to production standards. Participant P3 noted: “The componentization of this code (generated by DesignCoder) is better, allowing me to reuse it in multiple places.” This advantage is further reflected in code availability (DesignCoder: 4.52 vs. Baseline: 3.57), where modifications to a component generated by DesignCoder automatically propagate to all instances where it is used. Finally, we asked participants whether they would consider using DesignCoder. Most responded positively, with comments such as: \textit{“I would give it a try. It could save me a lot of time for complex UI layouts.”}

\begin{table*}[thp]
\centering
\caption{Participant ratings of DesignCoder vs. Baseline across different metrics.}
\resizebox{0.7\linewidth}{!}{
\begin{tabular}{@{}lccccc@{}}
\toprule[1.2pt]
\textbf{Method}              & \textbf{Code Availability} & \textbf{Modification Time} & \textbf{Readability} & \textbf{Maintainability} \\ \midrule
DeclarUI      & 3.57              & 3.24              & 4.43        & 3.32            \\
DesignCoder       & 4.52              & 4.20              & 4.75        & 4.32            \\
P-value         & 2.10E-6              & 4.57E-5              & 4.96E-2        & 1.51E-6            \\
\bottomrule[1.2pt]
\end{tabular}}
\label{tab:human evaluation}
\end{table*}

\subsubsection{Case Study}

\begin{figure*}[thp]
\centering
\includegraphics[width=0.8\textwidth]
{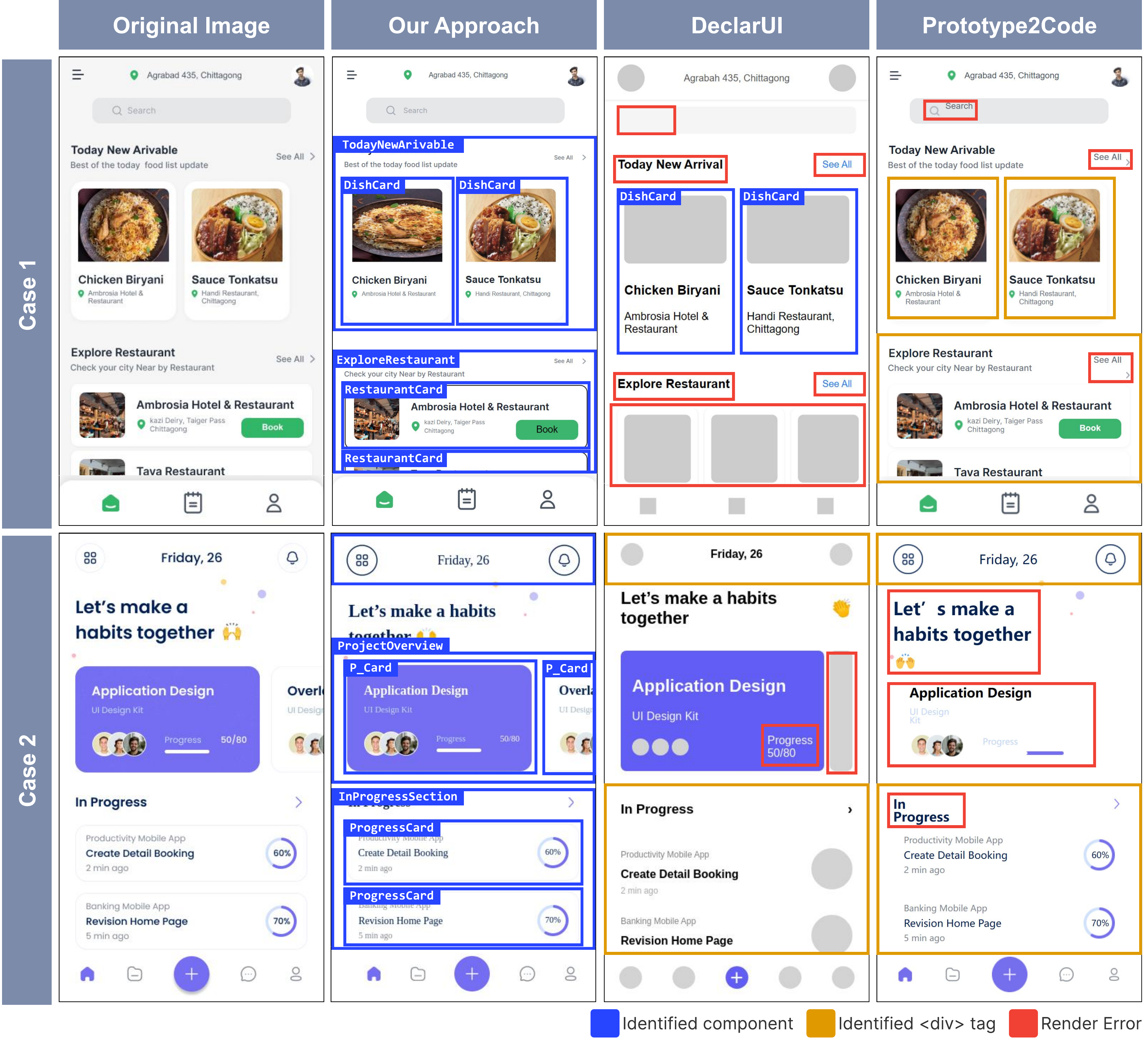}
\vspace{-0.1in}
\caption{Examples illustrating the effectiveness of DesignCoder in rendering and code structure, compared with two baseline methods.}
\label{fig:case1}
\vspace{-0.1in}
\end{figure*}
\begin{figure*}[thp]
\centering
\includegraphics[width=0.6\textwidth]{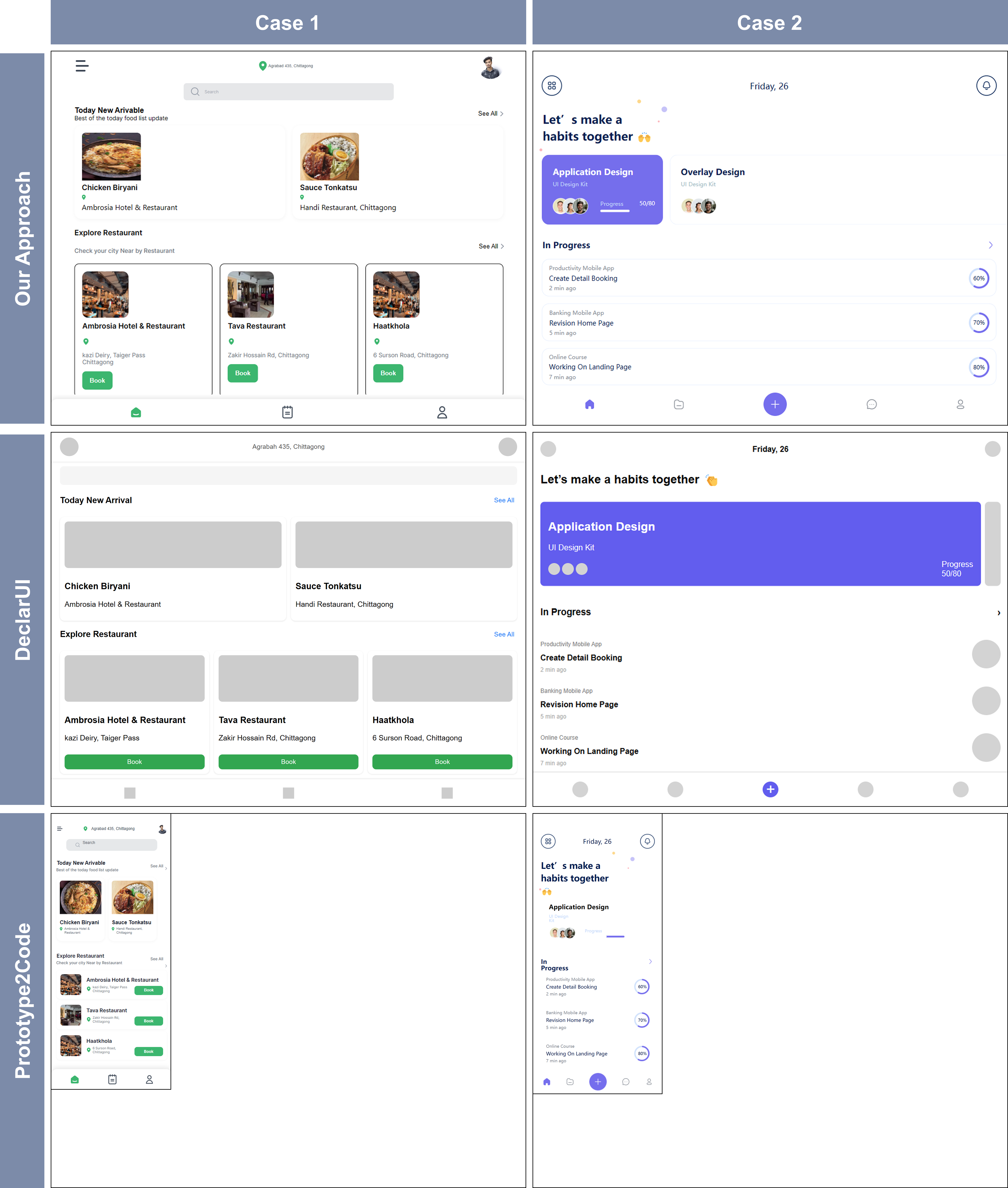}
\vspace{-0.1in}
\caption{Examples illustrating DesignCoder's responsiveness across different screen sizes, compared with two baseline methods.}
\label{fig:case2}
\vspace{-0.1in}
\end{figure*}

To demonstrate the superiority of DesignCoder, we provide specific examples in Figure \ref{fig:case1} and Figure \ref{fig:case2}. We selected two baseline methods, DeclarUI \cite{zhou2024bridging} and Prototype2Code \cite{xiao2024prototype2code}, for comparison. DeclarUI represents the SOTA approach that generates frontend code from UI screenshots, while Prototype2Code is the SOTA approach that generates frontend code from UI mockups. Our analysis reveals that DesignCoder outperforms these baselines in three key aspects: visual fidelity, component semantic completeness, and responsive compatibility. We illustrate these aspects using design files sourced from the Figma community and the company. 

Figure \ref{fig:case1} illustrates that DesignCoder achieves superior visual fidelity by accurately reconstructing component layouts and styles. The red boxes highlight visual inconsistencies in DeclarUI and Prototype2Code. Specifically, DeclarUI exhibits missing components (e.g., icons and background images) and misaligned text. While Prototype2Code maintains a relatively high level of visual accuracy, it still suffers from misplaced text and missing background elements.

In terms of component integrity, our approach effectively identifies design patterns and reuses identical components, ensuring clean and maintainable code. The blue boxes mark well-structured components in the code, while the yellow boxes highlight improper implementations that rely solely on ``<div>'' elements. We observe that DeclarUI correctly detects standalone ``Card'' components; however, it struggles to recognize Sections and List Items composed of multiple components, due to limitations in its computer vision model. Prototype2Code, on the other hand, fails to recognize components entirely, as it follows a rule-based approach for DOM tree generation, neglecting element semantics. By contrast, our method leverages the strong semantic understanding capabilities of MLLM. Through a multi-stage generation process, our approach ensures the correct implementation of complex components. 

Another key advantage of our method is its robust support for responsive design, as shown in Figure \ref{fig:case2}. It maintains optimal rendering across different window sizes by accurately determining code hierarchy, allowing precise calculation of relative sizes and positions for seamless cross-device adaptation. While DeclarUI demonstrates a certain degree of responsiveness, its generated style code is incomplete, leading to suboptimal rendering results. Prototype2Code, despite achieving high visual fidelity at the original page size, lacks support for multi-screen adaptation. In contrast, our approach strikes a balance between high-fidelity rendering and responsive behavior across different screen sizes, making it a practically viable solution for industrial applications.

\section{Threats to Validity}

There are two main threats to the validity of our approach. The threat to internal validity lies in the reliability of evaluation metrics. CLIP Score and SSIM, commonly used for UI visual similarity assessment, may fail to capture subtle differences. To address this, we introduce MSE to compare images at the pixel level. However, complete visual similarity does not necessarily indicate high-quality generated code. Therefore, we incorporate code structure similarity metrics. Nevertheless, manually annotated code structures may not fully reflect best practices in actual code implementation. As a result, the robustness of this approach still requires further validation in future work.

The threat to external validity involves variations in MLLM performance and dataset scale. Different MLLMs may produce different results depending on the prompt used, meaning our statistical findings may not fully reflect the optimal code generation performance of MLLMs. Additionally, although we selected two representative data sources and used 300 samples, our dataset size may not cover all variations of mobile application UIs. To mitigate this issue, future research should expand both the scale and diversity of UI samples to further validate the generalizability of the approach.
\section{Related Work}

\subsection{Image-to-code Generation}
In recent years, many studies have focused on generating UI code from images. Existing approaches can be categorized into two main types: deep learning-based \cite{beltramelli2018pix2code,robinson2019sketch2code,chen2022code, chen2018ui,xu2021image2emmet}, and MLLM-based approaches \cite{si2024design2code,zhou2024bridging,gui2025webcode2m,gui2025uicopilot,wan2024automatically}. Pix2Code \cite{beltramelli2018pix2code} was the first to employ deep learning techniques to generate code from GUI screenshots. Another notable computer vision-based method is Sketch2Code \cite{robinson2019sketch2code}, which converts hand-drawn sketches into HTML. To generate complex nested UI structures, Wu et al. \cite{wu2021screen} introduced the screen parsing problem, leveraging Faster R-CNN \cite{ren2016faster} to encode UI screenshots and integrating an LSTM model to generate UI hierarchy graphs. However, this approach remains limited in handling highly complex UI hierarchies with deeply nested structures.

\subsection{Mockup-to-code Generation}
In industrial front-end development, designers create UI mockups, while front-end engineers translate these mockups into code. However, this manual process is often time-consuming and prone to errors. To streamline development, mockup-to-code techniques have emerged, aiming to automate UI code generation \cite{linfigma2code, xiao2024prototype2code, codia}. Compared to visual-based code generation, approaches that integrate design metadata produce more structured and maintainable code. This multimodal method better supports industrial front-end development by enhancing consistency and efficiency. For example, Figma2Code \cite{linfigma2code} employs a metadata-based annotation approach to identify GUI components for code generation. However, due to the lack of UI component type recognition, the generated code often fails to meet development requirements. Enterprise-level platforms such as Imgcook \cite{Imgcook} and CodeFun \cite{Codefun} leverage vision models to identify components and layout structures, converting the results into a generic domain-specific language. They then apply various rules to generate front-end code across frameworks. While these platforms are highly efficient in meeting large-scale demands, they face persistent challenges such as UI element fragmentation and perceptual grouping, necessitating further manual adjustments to the generated code. Prototype2Code \cite{xiao2024prototype2code} enhances UI mockups by incorporating design lint techniques to refine the input, enabling high-fidelity UI page generation. However, the generated code lacks component semantics, resulting in an overuse of <div> elements rather than structured, meaningful component hierarchies. This study focuses on generating structured, component code and consistent visual styles. By integrating automated program repair techniques, our approach aims to enhance code usability, maintainability, and adherence to front-end development best practices.

\section{Conclusion}

In this paper, we propose DesignCoder, a novel hierarchy-aware and vision-guided self-correcting approach for generating high-quality UI code from design mockups. DesignCoder enhances MLLMs' UI understanding through a multimodal chain-of-thought reasoning framework, effectively recognizing complex nested structures. Empirical evaluations demonstrate that DesignCoder outperforms both state-of-the-art baselines and the latest multimodal large language models. Our approach bridges the gap between MLLMs and high-fidelity front-end development, making MLLM-generated UI code practically viable for industrial applications. It accelerates product iteration cycles, allowing development teams to focus more on core functionalities and product innovation. In future work, we plan to incorporate explicit design intent and leveraging component libraries to further enhance dynamic component behaviors and context-aware state management.
\bibliographystyle{ACM-Reference-Format}
\bibliography{bibliography}

\end{document}